\documentclass[preprint,showpacs,preprintnumbers,amsmath,amssymb]{revtex4}

\usepackage{graphicx}
\usepackage{dcolumn}
\usepackage{bm}

\newcommand\mse{Department of Material Science and Engineering, Nanjing University, Nanjing 210093, P. R. China }
\newcommand\ccmp{Group of Computational Condensed Matter Physics,
                 National Laboratory of Solid State Microstructures, Nanjing University, Nanjing 210093, P. R. China }
\newcommand\phy{Department of Physics, Nanjing University, Nanjing 210093, P. R. China }
\newcommand\shaoxing{Department of Physics, Shaoxing University, Shaoxing 312000, P. R. China}

\begin{document}

\title{Enhanced spin-orbit coupling in hydrogenated and fluorinated graphenes studied from first principles}

\author{Jian Zhou$^{1,3}$}
\author{Qifeng Liang$^{4}$}
\author{Jinming Dong$^{2,3}$}\thanks{Corresponding author: jdong@nju.edu.cn}
\affiliation{
$^1$ \mse \\
$^2$ \phy \\
$^3$ \ccmp \\
$^4$ \shaoxing }

\date{\today}




\begin{abstract}

  The spin-orbit couplings (SOCs) of hydrogenated and fluorinated graphenes are calculated from the first principles
  method. It is found that the SOC-induced band splittings near their Fermi energies can be significantly enhanced to
  the order of 10$^{-2}$ eV from the original about 10$^{-6}$ eV of the pure graphene, which is comparable to those
  found in the diamond and even the archetypal semiconductors. And two different mechanisms are proposed to explain
  the SOC enhancements in these two systems. The huge SOC enhancements are found to come not only from the sp$^3$
  hybridization of carbon atoms, but also from the larger intrinsic SOC of the fluorine atom than the carbon one.
  We hope many interesting phenomena caused by the SOCs (e.g. the spin Hall effect) can be observed
  experimentally in these systems.

\end{abstract}

\maketitle

\section{INTRODUCTION}

  The graphene, a single layer graphite, has attracted much attentions of researchers since its
  discovery in 2004 ~\cite{novoselov} due to its unique two-dimensional (2D) geometrical structure and
  Dirac-Fermion's electronic properties ~\cite{rmp}. Many works focus on its transport properties since
  the charge carrier mobilities of the graphene can reach to more than
  10$^4$ cm$^2$V$^{-1}$s$^{-1}$~\cite{novoselov,geim,zhang}.
  In addition, the graphene might also be a promising material for spin transport~\cite{tombros} and other
  spintronics applications, such as the spin-qubits~\cite{Trauzettel}, due to the weak spin-orbit coupling (SOC)
  of carbon atom. This is true since many theoretical works show that the SOC-induced band gap at the $K$ point
  (i.e. Dirac point) of the graphene is very small, which is at the order of 10$^{-6}$ $\sim$ 10$^{-5}$
  eV~\cite{min,hernando,yao,boettger}.

  However, although the weak SOC is good for transporting spin information, but it is probably not favored by others.
  As we known, the SOC is related to many interesting phenomena, such as optical dichroism, anomalous Hall effect,
  and spin Hall effect, among which the spin Hall effect especially offers a possibility of pure electrically driven
  spintronics in semiconductors and simple metals ~\cite{guo,yao2}. Unfortunately, the tiny SOC in a pure graphene makes
  impossible to observe experimentally the proposed spin Hall effect even the temperature goes down to
  0.01 K~\cite{min,hernando,yao}.

  Therefore, it is quite meaningful and important to study how to enhance the SOC of the
  graphene for other applications. In fact, the curvature effect can mix the $\pi$ and $\sigma$ electrons in the
  graphene and enhance the SOC significantly ~\cite{hernando}, which has been experimentally found in the single-walled
  carbon nanotube (SWNT)~\cite{exp}. Our recent first principles calculation shows that the SOC-induced band splitting
  can reach to several meV in small diameter SWNTs ~\cite{zhou}. Another possible method to enhance the graphene's SOC
  is to form sp$^3$ bonds by the impurities adsorbed on the sp$^2$ graphene, which has been proposed by Neto and Guinea
  very recently~\cite{neto}. They show that the impurities can induce the sp$^3$ distortion of the flat graphene, leading to
  a large SOC enhancement in it, which possibly reaches to about 10 meV, the same order as the intrinsic SOC in the
  carbon atom.

  On the other hand, a new kind of 2D crystal, called graphane, has been firstly predicted theoretically~\cite{sofo}
  and then synthesized recently in the experiment~\cite{elias}. The graphane is nothing but a fully hydrogenated
  graphene (H-graphene), in which the hydrogen atoms change the hybridizations of carbon atoms from sp$^2$ to sp$^3$, removing
  the conducting $\pi$ bands near the Fermi energy of the pure graphene. As a result, the H-graphene is a wide gap
  semiconductor, which can also be considered as a 2D-like diamond. In fact, even far before the H-graphene is found, the
  fluorinated graphite (F-graphite) has been synthesized and calculated~\cite{charlier,kurmaev,takagi}.
  It is then straightforward that the fluorinated graphene (F-graphene) can also exist and be easily synthesized just like
  the F-graphite. The F-graphene has the similar sp$^3$ hybridizations as the H-graphene and is also a wide gap
  semiconductor. However, since fluorine has a much larger electronegativity than hydrogen, we expect that in the
  F-graphene, the charge transfer between the graphene and dopants is quite different from that in the H-graphene.  And the
  SOC in a fluorine atom itself is also different from those in the hydrogen and carbon atoms.

  Based on the above arguments, we have used the first-principles method to calculate the practical sizes of SOC in the
  H-graphene and F-graphene. The paper is organized as follows. In Sec. II, the
  computational method and details are described. The main numerical results and discussions are given in Sec. III.
  Finally, in Sec. IV, a conclusion is presented.

\section{COMPUTATIONAL METHODS}

   The spin-orbit couplings in the three different systems, i.e., the pure graphene, H-graphene and F-graphene have been
   calculated by the density functional theory in the local density approximation implemented in the
   VASP~\cite{kresse1,kresse2} code, in which the projected augmented wave (PAW) method~\cite{paw2,paw1} and the
   Ceperley-Alder-type exchange-correlation are used. The plane-wave cutoff energy is 520 eV throughout the calculations.
   A supercell is used to simulate the two dimensional structures, in which the closest distance between two
   nearest layers is taken to be at least 10 \AA. A uniform grid of $20 \times 20 \times 1$ $k$ points is used for pure
   graphene and $10 \times 10 \times 1$ $k$ points for H-graphene and F-graphene. Both of the atomic positions and the
   in-plane lattice constant are firstly optimized without SOC and the residual forces at all atoms are less than
   0.005 eV/\AA. Once the optimized structure is obtained, the band structures with and without the SOC are calculated.
   By comparing the two band structures, the SOC-induced band splitting can be obtained.

   We use the same 'chair' structure in the H-graphene as that in the reference~\cite{sofo}, whose binding energy is about
   0.06 eV/atom lower than the 'boat' structure.
   The F-graphene is constructed by replacing the hydrogen atoms in the H-graphene with fluorine ones.
   As a reference, the band structure and SOC of the pure graphene is also calculated.

\section{ RESULTS AND DISCUSSIONS }

  \begin{table}
    \caption{\label{tab:8x} The calculated bonding length, bond angles, energy gap, and SOC-induced band splitting
                      near the Fermi energy in the pure graphene, H-graphene and F-graphene. The value of diamond is
                      the experimental one.}
    \begin{ruledtabular}
    \begin{tabular}{cccccc}
             & graphene  &  H-graphene  & F-graphene  & diamond \\ \hline
             C-C (\AA) & 1.413 & 1.516 & 1.553  &  1.544 \\
             C-H(F) (\AA) & -     & 1.117 & 1.365  &  -\\
             $\angle$ C-C-C ($^\circ$) & 120.0 & 111.6 & 110.6 & 109.5 \\
             $\angle$ C-C-H(F) ($^\circ$) & -     & 107.3 & 108.3 & - \\
             Energy gap (eV) & 0 & 3.38 & 2.96  & 5.48 \\
             SOC size (meV) & 10$^{-3}$ & 8.7 & 27.4  &  6.0
    \end{tabular}
    \end{ruledtabular}
   \end{table}

   \begin{figure}
      \includegraphics{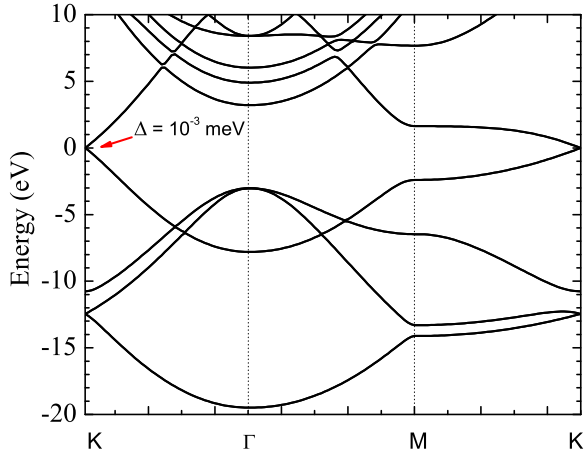}
      \caption{\label{fig:1} (Color online) The band structure of pure graphene with SOC.}
    \end{figure}

 \subsection{Pure graphene}

    The pure graphene is well known to be a zero gap semiconductor, whose band structure is given in Fig.1. The linear
    band structures of $\pi$ electrons touch exactly on the Fermi energy at the $K$ point. When the SOC is turned on, a
    small gap can be opened at this point, which is at the order of 10$^{-6}$ eV. This is because in the pure graphene
    only the $\pi$ electrons hopping between the two next nearest neighbor carbon atoms can contribute to the SOC, which
    is the second order process and quite tiny~\cite{min,hernando,yao}. However, at the $\Gamma$ point, the two-folding
    degenerate valence band splits by about 9.4 meV, which is caused by the $\sigma$-electrons and much larger than that
    at the $K$ point.

    Since only the electrons near the Fermi energy participate the transport process, some spin-related phenomena cannot
    be observed until the temperature is as low as 0.01 K. In order to get a large SOC at the Fermi energy, we have to use
    the $\sigma$ electrons. The recently discovered H-graphene gives us such a system, where the $\pi$ bands are removed by
    the hydrogen atoms and the $\sigma$ bands become the highest occupied molecular state (HOMO) at the $\Gamma$ point.
    We expect a large SOC can be found in the H-graphene.

 \subsection{H-graphene}

   \begin{figure}
      \includegraphics{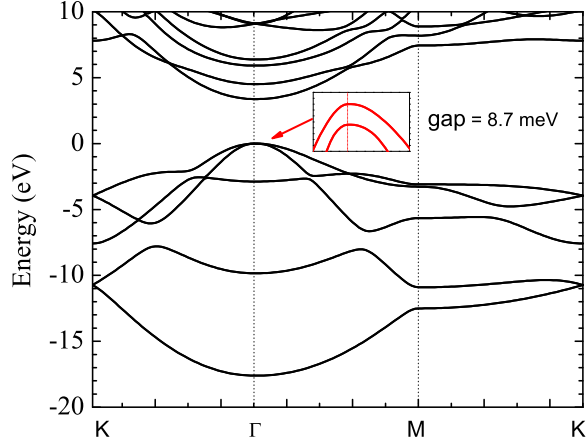}
      \caption{\label{fig:2} (Color online) The band structure of H-graphene with SOC. The inset
           shows the detailed band structure near the $\Gamma$ point.}
    \end{figure}

    First of all, we can see from Table I that the bond lengthes of C-H and C-C are 1.117 and 1.516 \AA,
    respectively, which are well consistent with Sofo \emph{et al.}'s results ~\cite{sofo}. The angle between the C-C and
    C-H bonds is about 107.3$^\circ$ and that between the two C-C bonds is about 111.6$^\circ$. All these bond lengthes
    and angles are much like the ones in the diamond rather than in the pure graphene, indicating that the H-graphene is a
    pure sp$^3$ carbon system and is somewhat like to a flat 2D diamond. The band structure of chair-type H-graphene is
    given in Fig. 2, from which it can be clearly seen that the H-graphene is a wide direct gap semiconductor with its band
    gap of about 3.38 eV, which is underestimated due to the well-known DFT band gap problem. A recent GW calculation
    shows that the band gap of H-graphene is 5.4 eV, which is very close to that of the diamond~\cite{gw}.

    It is clearly shown by a comparison of Fig.1 with Fig.2 that the $\pi$ band in the pure graphene disappears in the
    H-graphene and the $\sigma$ band at the $\Gamma$ point becomes the top of its valence band. When considering SOC,
    the two fold degenerate $\sigma$-bands split near the Fermi energy at the $\Gamma$ point with a gap of 8.7 meV,
    as shown in the inset of Fig.2. This value is at the same order as the intrinsic SOC of atomic carbon and greatly
    enhanced, compared with the negligible SOC near the Fermi level of the pure graphene, which is induced by the
    $\pi$-bands in it. This SOC-induced band splitting appears at the highest point of the valence band and is very
    similar to the SOC in the archetypical semiconductors ~\cite{guo}.

\subsection{F-Graphene}

   \begin{figure}
      \includegraphics{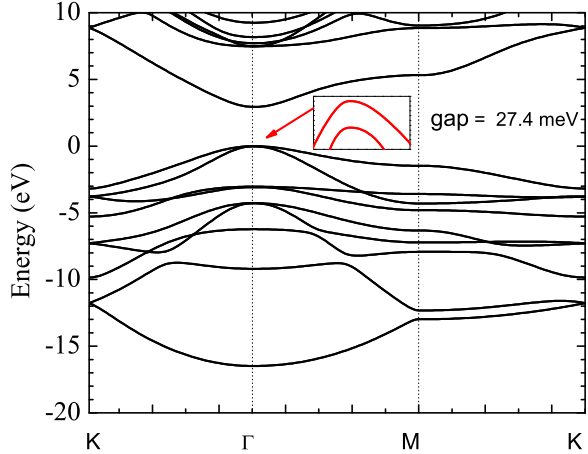}
      \caption{\label{fig:3} (Color online) The band structure of F-graphene with SOC. The inset
           shows the detailed band structure near the $\Gamma$ point.}
    \end{figure}

    The fluorinated graphene is also calculated and its band structure is given in Fig.3. Obviously, the F-graphene
    has the similar band structure with the H-graphene, except that more valence bands are found in the F-graphene.
    The F-graphene is also a wide direct gap semiconductor with a band gap of 2.96 eV, a little smaller than that of the
    H-graphene. However, we also expect that the real band gap is about 5 eV if we take into account the same DFT
    underestimation of the band gap as in the H-graphene. This also gives us a method to tune the band gap of pure graphene
    by using different adsorbed atoms.

    In Table I, the C-C and C-F bond lengthes in the F-graphene are shown to be 1.553 and 1.365
    \AA, respectively. The angle between the C-C and C-F bonds is about 108.3$^\circ$, and that between the two C-C bonds
    is about 110.6$^\circ$, which are all well consistent with the previous calculation ~\cite{charlier}.
    Although the bond lengthes are much longer than that in the H-graphene, the F-graphene is also a sp$^3$ carbon
    system and therefore both of them have the similar band structures.

    It is the most interesting to find that the SOC-induced band splitting in the F-graphene can reach to 27.4 meV near the
    Fermi energy, which is far bigger than that of carbon atom. We believe that this large SOC cannot be explained only by the
    sp$^3$ hybridization of carbon atoms. What is the difference between the H-graphene and F-graphene? We think the main
    difference between them is that the fluorine atom has a much larger electronegativity than hydrogen atom. The charge
    transfers from the carbon atom to fluorine one in the F-graphene and the SOC splitting is contributed from both the
    carbon and fluorine atoms. In order to confirm this hypothesis, we have calculated the site and orbital projected
    wavefunction of HOMO at the $\Gamma$ point, whose coefficient can give the band characteristics. The obtained results
    show that the HOMO state of the F-graphene consists of about 60\% carbon's $p$-electron and about 40\%
    fluorine's $p$-electron. It is known that the intrinsic SOC of the fluorine atom is quite larger, which is about 50
    or 100 meV by different experiments~\cite{F}. So, the large SOC splitting in the HOMO state of the F-graphene comes
    mostly from that of the fluorine atom, whose mechanism is totally different from that in the H-graphene, where its HOMO state
    consists of almost 100\% carbon's $p$-electron. As a result, the SOC splitting near the Fermi energy in the H-graphene has
    the similar value as that of diamond or carbon atom, while the one in the F-graphene has a much bigger value.

\section{ CONCLUSION }

   In summary, we have used the first-principles method to calculate the practical SOC-induced band splittings of the
   H-graphene and F-graphene. And two different mechanisms have been
   proposed to explain the great enhancements of the SOC splitting in both of them, in which the carbon atoms'
   hybridizations become the sp$^3$ from their original sp$^2$ due to the doped hydrogen or fluorine atoms, making the
   $\sigma$-bands move to the top of their valence bands and a large energy gap to be opened at there.
   The SOC induced band splitting near the Fermi energy of the H-graphene is found to reach to 8.7 meV, which is
   significantly enhanced by four orders from its original value near the Fermi energy of the pure graphene.
   More importantly, it is even enhanced to 27.4 meV in the F-graphene, which mainly
   comes from the larger SOC of fluorine atom than the carbon atom. These large SOC values are comparable, respectively,
   to those found in the diamond and archetypal semiconductors. So, we hope that the spin Hall effect and other
   SOC-related phenomena can be easily found in the H-graphene and F-graphene than that in the pure graphene.

\begin{acknowledgments}

  This work was supported by the Natural Science Foundation of China under Grant No. 10874067, and also from a Grant
  for State Key Program of China through Grant Nos. 2004CB619004 and 2006CB921803. J.Z. is also supported by the China
  Postdoctoral Science Foundation funded project and the Jiangsu Planned Projects for Postdoctoral Research Funds.

\end{acknowledgments}

\end{document}